\documentclass[reprint,english,aps,prl,showpacs,superscriptaddress,preprintnumbers]{revtex4-1}
\usepackage[T1]{fontenc}
\usepackage[latin9]{inputenc}
\usepackage{float}
\usepackage{amsmath}
\usepackage{amssymb}
\usepackage{graphicx}
\usepackage{xcolor}

\makeatletter
\@ifundefined{textcolor}{}
{%
 \definecolor{BLACK}{gray}{0}
 \definecolor{WHITE}{gray}{1}
 \definecolor{RED}{rgb}{1,0,0}
 \definecolor{GREEN}{rgb}{0,1,0}
 \definecolor{BLUE}{rgb}{0,0,1}
 \definecolor{CYAN}{cmyk}{1,0,0,0}
 \definecolor{MAGENTA}{cmyk}{0,1,0,0}
 \definecolor{YELLOW}{cmyk}{0,0,1,0}
}

\usepackage{amstext}
\usepackage{babel}

\newcommand{\comment}[1]{}
\addto\captionsenglish{}
\makeatother

\begin{document}
\title{Tunable narrow-band plasmonic resonances in electromagnetically-induced-transparency media}
\author{D. Ziemkiewicz}
\affiliation{Institute of Mathematics and Physics, UTP University of Science and Technology, Al. Prof. S. Kaliskiego 7, 85-789 Bydgoszcz, Poland}
\author{K. S\l owik}
\email{karolina@fizyka.umk.pl}
\affiliation{Institute of Physics, Faculty of Physics, Astronomy and Informatics, Nicolaus Copernicus University, Grudziadzka 5, 87-100 Torun, Poland}
\author{S. Zieli\'{n}ska - Raczy\'{n}ska}
\affiliation{Institute of Mathematics and Physics, UTP University of Science and Technology, Al. Prof. S. Kaliskiego 7, 85-789 Bydgoszcz, Poland}

\begin{abstract}
The spectral response of a plasmonic nanostructure may heavily depend on the refractive index of its surroundings. 
The key idea of this paper is to control this response by coherent optical means, i.e. with an optically controlled electromagnetically-induced-transparency medium.
In such environment, an external laser provides a knob to shift the position of plasmonic resonances without the need to change the geometry of the nanostructure. 
Additionally, the setup can be exploited to excite narrow-band surface plasmon polaritons. 
\end{abstract}


\maketitle
\section{Introduction} \label{sec:introduction}
Plasmonic nanostructures offer an unprecedented potential to control and to tailor spectrally sensitively the distributions of electromagnetic fields \cite{Bryant}.
They show extraordinarily strong confinement of electromagnetic energy into nanoscopic regions of space,
and the corresponding huge enhancement of its local density \cite{Kuttge}.
They can be exploited to force the desired propagation directions of light \cite{Bonod},
to mediate between quantum systems and their macroscopic surroundings \cite{Akselrod},
to boost the sensitivity of sensors up to a single-molecule level \cite{Liu2011},
to increase by orders of magnitude the efficiency of solar cells \cite{Esteban},
eventually to improve the efficiency of cancer therapy~\cite{Hirsch}.

Plasmonic devices, however, suffer from a severe disadvantage which relates to the inability to dynamically tune the properties of metallic nanostructures. 
Their optical response can be tailored by carefully choosing their geometry, and their dielectric surroundings.
In other words, once a nanostructure is fabricated, its optical properties are fixed to a large extent. Since 2004, there is a considerable interest in the field of so-called active plasmonics \cite{Krasavin}, where various means to control surface plasmon polariton (SPP) propagation are sought.
Previous attempts to achieve tunability of nanodevices were mainly focused on temperature control \cite{Kaplan,Nikolajsen,Lereu}, which is relatively slow and excludes dynamic modulation. A faster way is achieved with a modification of charge distribution \cite{MacDonald}, or through integration of liquid crystals \cite{Si}. A different approach based on phase change materials allowed to achieve tunable radiation patterns and was explored in Ref.~\cite{Alaee}.
We propose to exploit the electromagnetically induced transparency (EIT) phenomenon to achieve fast optical tunability of SPPs by dynamically adjusting an external, continuous laser illumination. EIT has been observed in evanescent fields present at the reflection on the medium interfaces \cite{Thomas}.
\begin{figure}[ht!]
\begin{centering}
\includegraphics[width=6.6cm,keepaspectratio]{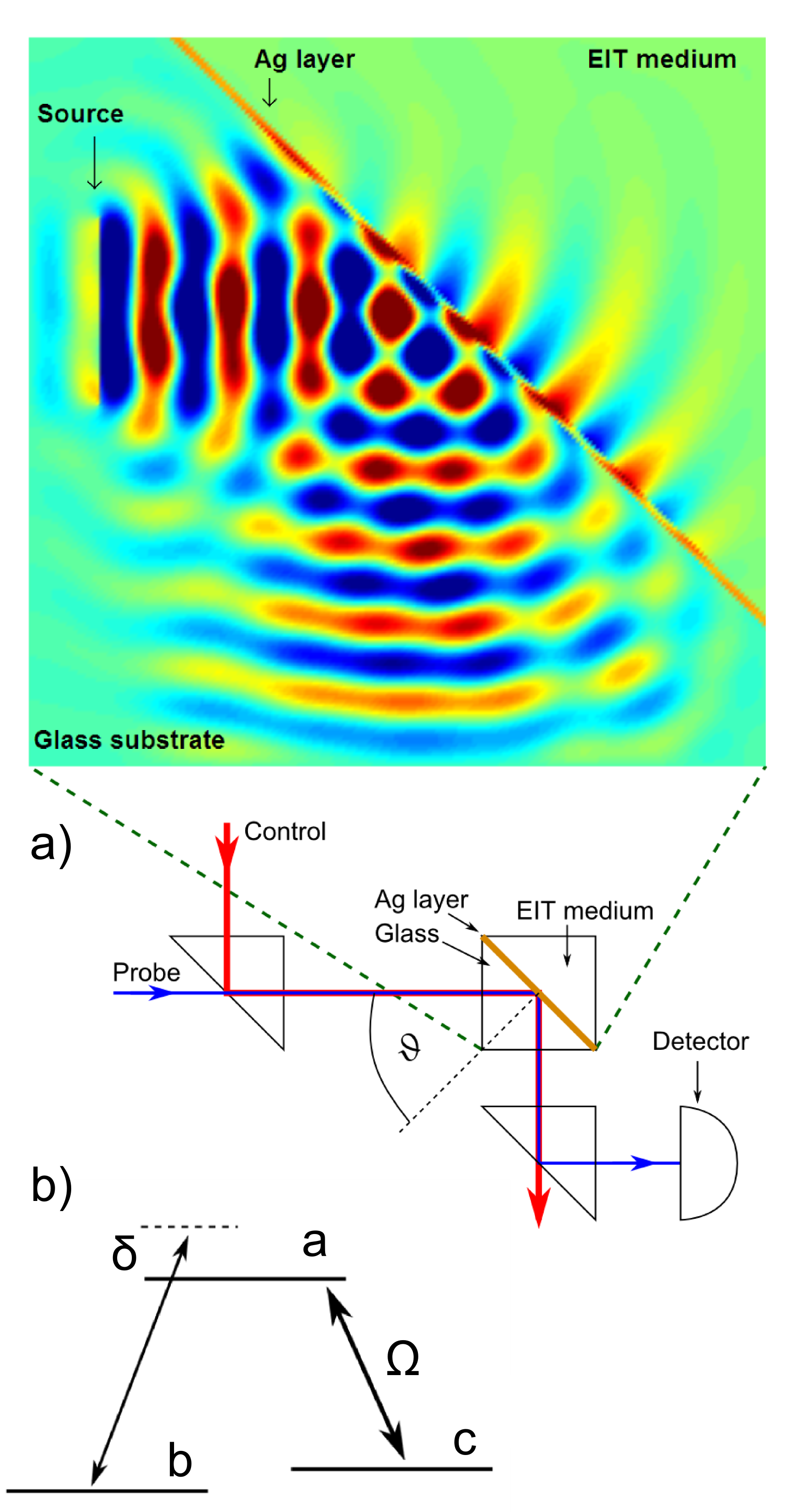}
\par
\end{centering}
\caption{\label{fig:setup} a) Scheme of the system under consideration and its implementation in FDTD simulation: a silver surface atop a glass substrate is covered with EIT-medium. Here, an incident beam hits the metal surface at $\vartheta=45^o$. b) A scheme of the atomic $\Lambda$ system. }
\end{figure} 
Another bottleneck of metal-based plasmonic devices relates to their extremely broad resonances.
Their exceptional width reflects huge scattering and absorption rates of plasmonic nanostructures,
that may be detrimental for certain applications.
For instance, when a coupling of nanoantennas to quantum emitters is considered, large resonance widths constitute a serious limitation on the types of effects that can be anticipated and observed \cite{Slowik}. 
Here we take advantage of the fact that if a metallic surface is placed in contact with a highly dispersive EIT medium, the SPPs will also exhibit sharp features. 

The above observations unlock the possibility to achieve SPP resonances that are both narrow and tunable. The idea to control reflection spectra of a beam illuminating thin metallic films with EIT was introduced in Ref.~\cite{Du2012}.
The Author examined the influence of gas parameters, namely density and additional gas admixture, of EIT media on the conditions of SPP creation. 
The concept was later developed to include the motion of atomic environment on reflection and transmission \cite{Du2015}.
Contrary, our contribution is focused at dynamic tunability of plasmonic excitations. 
The tunability can be achieved by optical means, through a modification of the environment of the metallic nanostructure.
We propose to exploit the EIT technique and apply it to steer the creation and to control the propagation of SPPs by dynamic adjusting the intensity of the control field. 

EIT \cite{Harris} is an important quantum-optical effect 
that allows for a coherent control of materials' optical properties. 
The generic EIT bases on extraordinary dispersive properties of an atomic medium with three active states in the $\Lambda$ configuration. 
This phenomenon leads to a significant reduction of absorption of a resonant weak probe laser field by irradiating the medium with a strong control field (coupling empty levels) that makes an otherwise opaque medium transparent. 
It causes dramatic changes of dispersion properties of the system: 
absorption forms a dip called a transparency window and approaches zero,
while dispersion at the vicinity of this region becomes normal with a slope, which increases for a decreasing control field. 
The resonant probe beam is now transmitted almost without losses. 
EIT has been explained by destructive quantum interference between different excitation pathways of the excited state
or alternatively in terms of a dark superposition of states. 
Since at least 20 years there has been a considerable level of activity devoted to EIT which has been motivated by recognition of a number of its applications, among which slowing and storing light are well-known examples \cite{Fleisch}. 

Extensive studies of EIT in atomic media inspired our interest in applications of this phenomenon to control and change on demand the spectral properties of  electromagnetic field tailored by nanoantenas.
Such control could be achieved by using the EIT-medium as the nanoantenna environment, on which the plasmonic response, 
e.g., the spectral position of plasmonic resonances, strongly depends.
Among the potential applications that would be unlocked with on-demand tuning, 
are nanodevices, based on metasurfaces or nanoantennas, that could be dynamicaly switched between their operational modes,
or tunable technologies for information processing both at a classical and at the quantum level.

The article begins with an introduction of the considered setup.
Next, its optical properties in terms of reflection coefficients, are derived and analyzed.
We find an excellent agreement of these analytical predictions with time-domain numerical simulations of Maxwell's equations, which is presented in the final part of the manuscript.

\section{Setup and operating principle} \label{sec:setup}
The setup that we consider consists of a silver nanosurface, 
spread on top of a glass substrate, and covered with a surrounding gas medium that supports EIT (Fig.~\ref{fig:setup} a). 
As an example we consider transitions in the sodium D2 line ($3^{2}S_{1/2}\rightarrow 3^{2}P_{3/2}$).
The upper state $a$ would simply be the $|3^{2}P_{3/2}, F=0, m_F=0\rangle$ hyperfine sublevel.
The lower states would correspond to symmetric and antisymmetric superpositions, respectively:
$b,c \sim |3^{2}S_{1/2}, F=1,m_F = -1\rangle \pm |3^{2}S_{1/2}, F=1,m_F = +1\rangle$.
This is necessary for them to be coupled with linearly polarized light, as required by our scenario:
the $c \leftrightarrow a$ transition is driven with TE-polarized light (the control field),
while the probe of TM polarization couples states $b$ and $a$ [Fig.~\ref{fig:setup} b)].

We consider two co-linear beams; the incident, weak probe beam of TM polarization illuminates the metallic surface from the glass side, creating SPPs at the metal - EIT medium interface, in a typical Kretschmann configuration. A stronger, TE polarized control beam generates coherences in EIT medium, altering its dispersive relation. In other words, it makes the medium transparent for the propagating SPP generated by the probe beam. The scenario can be realized when the following conditions are fulfilled: 
\begin{itemize}
\item the control beam should be resonant with $a$-$c$ transition to induce the transparency of the medium, 
\item the probe beam has to be resonant with the $a$-$b$ transition for it not to be absorbed by the EIT medium,
\item the angle of incidence $\vartheta$ of the probe beam is set to match the SPP resonance condition. 
\end{itemize}
The the first two requirements need not to be strictly satisfied: it is enough to set the detunings of the two beams equal, i.e. fulfill the two-photon resonance. If these conditions are met, an evanescent electromagnetic mode at the interface appears, which propagates along the surface: the probe beam can excite a propagating SPP (see Appendix). The surface plasmon resonance conditions depend on the incidence angle and frequency, silver layer thickness, 
and permittivity of the surrounding media (glass and EIT). To this end, the required minimal thickness of EIT layer is of the order of the optical wavelength $\lambda$ of the probe field \cite{Du2015}. In practice the coherence of the EIT medium is provided by the deeper penetration of the stronger, evanescent control field \cite{Fleisch, Thomas}. 
In the situation when a $\Lambda$ system is consituted by two hiperfine-splitted metastable states $b$ and $c$ 
the Doppler broadening has no adverse effect on EIT provided that one uses copropagating probe and control beams 
\cite{Zibrov,Fleisch}.
A practical way of SPP detection is the measurement of the reflection spectrum, as shown in Fig.~\ref{fig:setup} a). The resonance is manifested in a form of a pronounced dip in the spectrum.

The key idea of this paper is to apply the optical tunability of EIT to the field of plasmonics. In particular, we control the plasmon resonance condition through modification of the refraction coefficient of the surrounding EIT medium by adjusting the intensity of the control field. 
Surrounding characterized with a steep dispersion, due to its illumination by control field, will drastically modify the optimal conditions to excite SPPs. 
With such strong frequency dependence of the optical response of the medium, the spectral width of the excited plasmons will be comparable to the atomic linewidth. As it will be shown later,
the exploitation of EIT media additionally offers the tunability of such SPP's spectral position of unprecedented precision, achievable with a knob.  

In the next section, we describe the principles of the EIT phenomenon,
which are the basis of the results that follow.

\section{Electromagnetically induced transparency} \label{sec:eit}
As an electromagnetic beam, resonant with a certain atomic transition, propagates through a medium, it is normally absorbed and exponentially damped. 
This can be modified in special media described by a $\Lambda$ energy scheme [Fig.~\ref{fig:setup} b)],
with an initially fully occupied ground state $b$, an excited state $a$, and a side state $c$.
The transition frequencies between the $i$ and $j$ states will be denoted with $\omega_{ij}$.
If a relatively strong control field of Rabi frequency $\Omega$ couples the two empty states $a$ and $c$, 
the refractive index of the medium 
\begin{equation}\label{eq:n}
n_\mathrm{EIT}(\delta) = \sqrt{1+\mathrm{Re}\chi (\delta)}
\end{equation}
is modified, where instead of an optical frequency $\omega$ we write explicitly the dependence on 
a detuning from the transition resonance, defined as $\delta = \omega - \omega_{ab}$.
Here the Rabi frequency reads $\Omega = \frac{\mathbf{E}\cdot\mathbf{\mu}_{ac}}{\hbar}$, where $\mathbf{E}$ is the electric field of the control beam, and $\mathbf{\mu}_{ij}$ stands for the vector electric dipole moment of the $i\leftrightarrow j$ transition, and $\hbar$ denotes the reduced Planck constant. 
The crucial impact of the control field $\Omega$ can be explicitly seen from the electric susceptibility $\chi(\delta)$ of the $\Lambda$-medium
\begin{equation} \label{eq:chi}
 \chi(\delta) = -\frac{N|\mathbf{\mu}_{ab}|^2}{\hbar\epsilon_0}\frac{\delta-\delta_{ac}+i\gamma_{bc}}{(\delta+i\gamma_{ab})(\delta-\delta_{ac}+i\gamma_{bc})-|\Omega|^2},
\end{equation}
where $\delta_{ac} = \omega_c - \omega_{ac}$,
$\omega_c$ is the frequency of the control field,
$N$ stands for the density (number of atoms per unit volume) of the EIT medium, $\epsilon_0$ is the vacuum permittivity, and $\gamma_{ij}$ are decoherence rates between the corresponding states. The decoherence between the excited and lower states $\gamma_{ab}$ origins mostly from spontaneous emission, whose influence is much greater than of other factors such as collisions between atoms or with cell walls \cite{Fleisch}. Contrary, the dephasing between the lower states $\gamma_{bc}$ corresponds to an electric-dipole-forbidden transition, for which spontaneous emission is negligible. Therefore, the main source of $\gamma_{bc}$ are collisions of the medium atoms with the metallic film. 
All the important features of EIT remain observable even when $\gamma_{bc} \neq 0$, provided that the control field Rabi frequency satisfies $|\Omega|^2 \gg \gamma_{ab}\gamma_{bc}$ \cite{Fleisch}, which is the case in the following calculations.

In our model we set a typical value of density for EIT atomic vapor $N=10^{13}~\mathrm{cm}^{-3}$ \cite{Dziczek}. 
For an exemplary EIT medium, we choose sodium vapors at room temperature, whose D2 transition line is centered at $589.1$ nm \cite{Sod} (i.e. $3.198$ PHz).
The transition is characterized with a dipole moment \mbox{$\mu_{ab}= 1.72\times 10^{-29}$ C$^\cdot$m},
and decoherence rates \mbox{$\gamma_{ab} = 61.54$ MHz} \cite{Sod}, 
and typically \mbox{$\gamma_{bc} = 0.01\gamma_{ab}$}.

With the chosen set of parameters the susceptibility varies by $0.01$ which causes sufficient change of surrounding medium refraction index to achieve SPP control.

\begin{figure}[ht!]
\begin{centering}
\includegraphics[width=\linewidth]{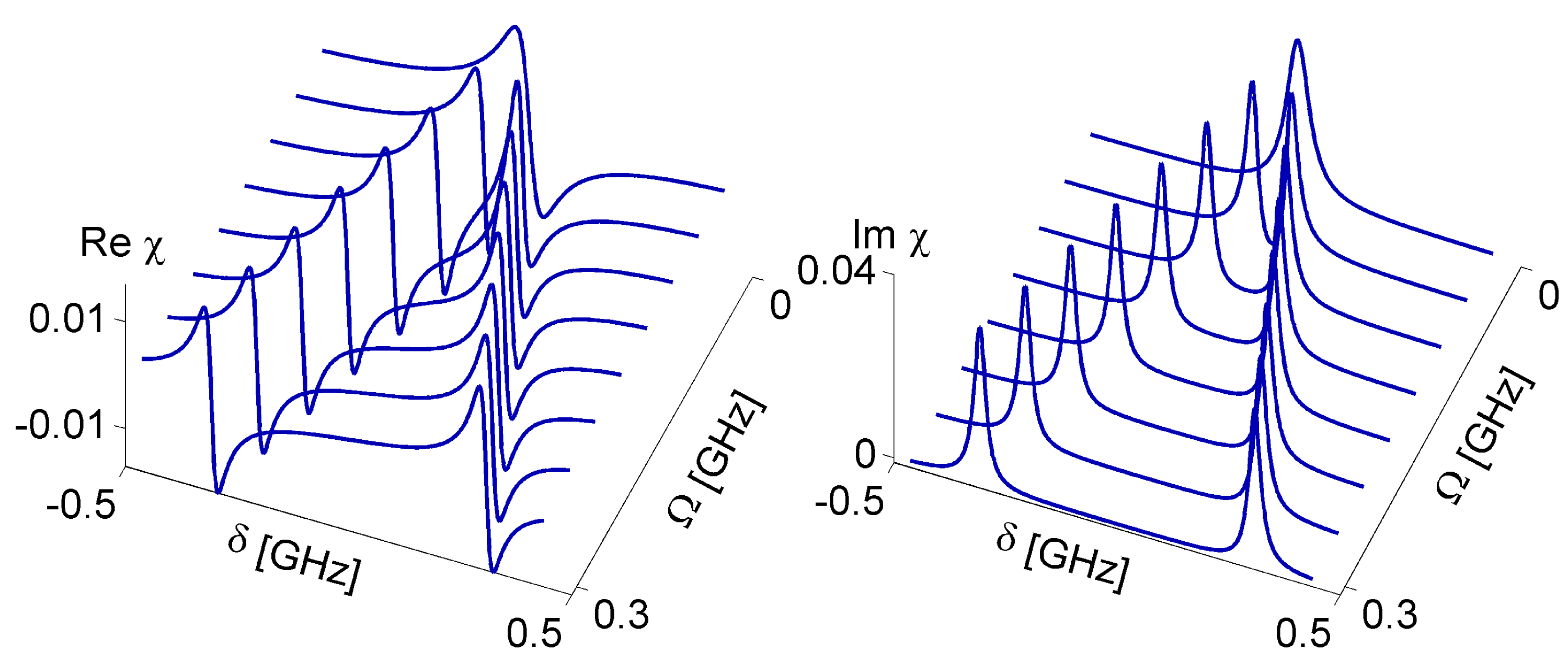}
\end{centering}
\caption{\label{fig:lambda} Real and imaginary parts of the electric susceptibility for a probe beam in a $\Lambda$ atomic medium, as a function of the probe detuning $\delta$ and the strength of the control beam in terms of the Rabi frequency $\Omega$. The control field detuning $\delta_{ac}=0$. }
\end{figure} 

The real and imaginary parts of the electric susceptibility, responsible for dispersion and absorption, are shown in Fig.~\ref{fig:lambda} for a set of control fields $\Omega$. 
At the absence of the control field $\Omega=0$, the absorption peak and normal dispersion correspond to an opaque medium. 
The control field opens a transparency window, with two absorption peaks at its sides at $\delta=\delta_{ac}\pm\Omega$. 
In other words, the width of the transparency window and the spectral position of the peaks can be tuned with the control field.
This observation lies at the heart of our contribution.

\section{Surface plasmon polaritons} \label{sec:polarit}
The resonance condition necessary for the SPP occurs when the polariton wave vector
\begin{equation}\label{eq:k}
k=\frac{\omega}{c}\sqrt{\frac{\epsilon_m\epsilon_s}{\epsilon_m+\epsilon_s}}
\end{equation}
matches the glass wave vector component parallel to the surface $k_{||}=\frac{n\omega}{c}\cos(\vartheta)$, where $\epsilon_m$ and $\epsilon_s$ are the permittivities of metal and surrounding, respectively, and $\vartheta$ stands for the incidence angle. In this manuscript, the index $s$ may refer to vacuum, glass or EIT medium. 
Therefore, once an EIT medium is used as surroundings for a plasmonic nanosurface, it modifies the SPP resonance condition. 
To a good approximation, such condition is fulfilled if the plasmon dispersion curve of the bulk metal meets the dispersion of the surroundings. 
Normally, the latter is chosen such that it does not show any transitions in the spectral region of interest and its refractive index $n_s\approx \mathrm{const}.$ 
\begin{figure}[ht!]
\begin{centering}
\includegraphics[width=\linewidth]{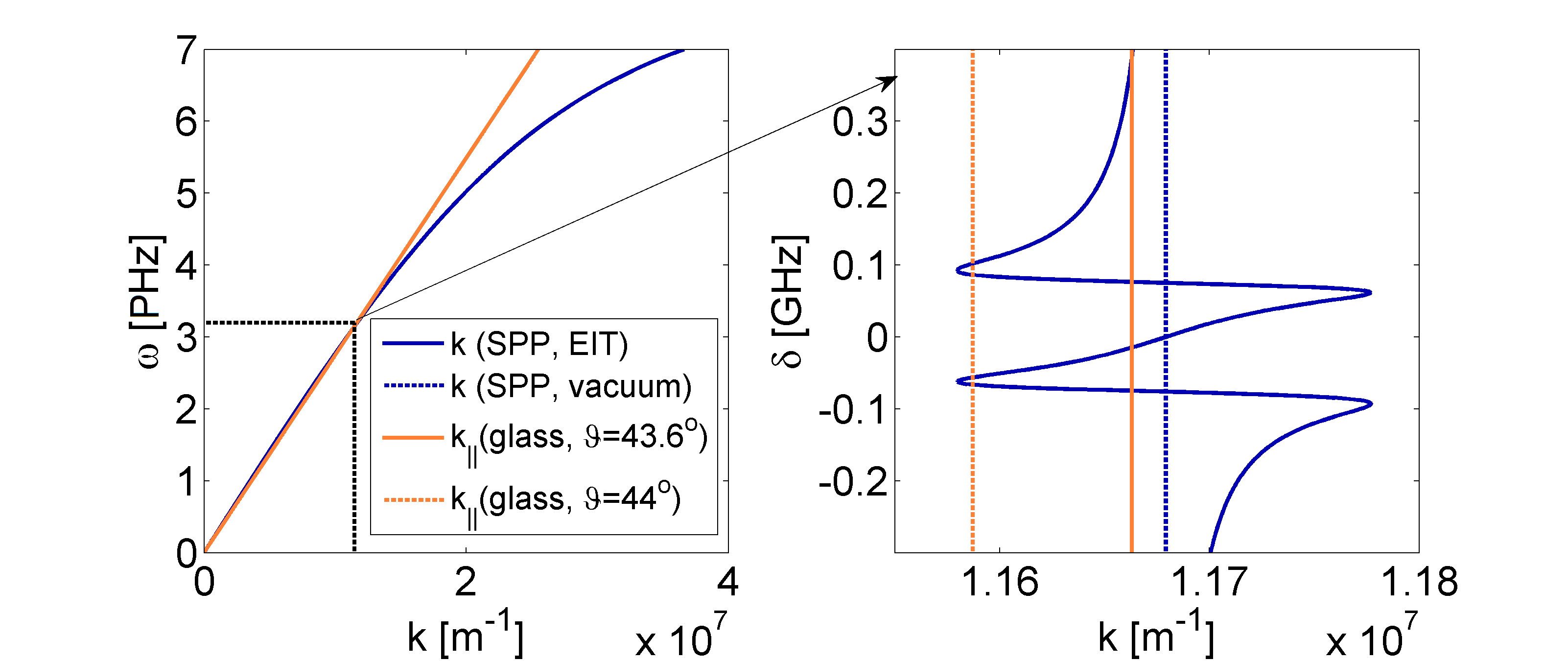}
\end{centering}
\caption{\label{fig:disp}Left: dispersion relation $\omega(k)$ for the metal-EIT medium plasmons (blue line) and the parallel wave vector component $k_{||}$ in glass. 
Right: the resonance region is enlarged, results for two incidence angles (solid and dashed lines) are shown. 
The crossing points correspond to SPP resonances.}
\end{figure} 
This means that the corresponding dispersion curve is given by a straight line $k=\frac{\omega}{cn_s}$,
neighbouring the metallic curve at a rather wide range, and resulting in a spectrally broad SPP resonance (see Fig~\ref{fig:disp}, where the dispersion relation for vacuum is marked with a dashed line, and is close to the glass dispersion curve for $\vartheta=44.6^o$ over the whole frequency range of interest). These curves correspond to the parameters (incidence angle, thickness of the silver layer) optimal to achieve the plasmonic resonance. On the other hand, if these optimal conditions are not met, an excitation of an SPP may be hindered, since the two dispersion curves are separated from each other. A strong spectral dependence of the atomic dispersion curve at the vicinity of a state transition, may either interrupt the broad plasmonic resonance range [Fig.~\ref{fig:disp}, orange line, $\delta \approx 0.1$ GHz] or restore resonance in the conditions where it normally would not be achieved [orange dashed line, $\delta \approx \pm 0.08$ GHz].
The existence of SPPs follows from characteristic field distribution; the corresponding pattern, obtained numerically, is shown in the Appendix. The SPPs are also manifested in the reflection spectra of the nanosurface. 
The latter can be derived analytically for the detection angle equal to the illumination angle $\vartheta$.
As follows from classical optics, the reflection coefficient at a combined interface reads \cite{Simon}
\begin{equation}\label{Rcoeff}
 R\left(\delta\right) = \left| \frac{r_\mathrm{m}+r_\mathrm{EIT}\left(\delta\right)e^{-2k\left(\delta\right)d}}{1+r_\mathrm{m}r_\mathrm{EIT}\left(\delta\right)e^{-2k\left(\delta\right)d}} \right|^2,
\end{equation}
where the reflection amplitudes at the glass-metal $r_\mathrm{m}$ and metal-EIT $r_\mathrm{EIT}$ interfaces read
\begin{eqnarray}
 r_\mathrm{m} &=& \frac{n_\mathrm{m} \cos \vartheta - n_\mathrm{g} \cos\vartheta_\mathrm{m}}{n_\mathrm{m} \cos \vartheta + n_\mathrm{g} \cos\vartheta_\mathrm{m}}, \\
 r_\mathrm{EIT}\left(\delta\right) &=& \frac{n_\mathrm{EIT}\left(\delta\right) \cos \vartheta_\mathrm{m} - n_\mathrm{m} \cos\vartheta_\mathrm{EIT}\left(\delta\right)}{n_\mathrm{EIT}\left(\delta\right) \cos \vartheta_\mathrm{m} + n_\mathrm{m} \cos\vartheta_\mathrm{EIT}\left(\delta\right)}.
\end{eqnarray}
The diffraction angles $\vartheta_\mathrm{m,EIT}$ in the metallic and EIT domains, can be directly derived from the Snell's law $n_\mathrm{g}\sin\vartheta = n_\mathrm{m}\sin\vartheta_\mathrm{m} = n_\mathrm{EIT}\left(\delta\right)\sin\vartheta_\mathrm{EIT}\left(\delta\right)$:
\begin{eqnarray}
 \cos \vartheta_\mathrm{m} &=&  \sqrt{1- \frac{n_\mathrm{g}^2}{n_\mathrm{m}^2}\sin\vartheta},\label{Angle_1}\\
 \cos \vartheta_\mathrm{EIT}\left(\delta\right) &=& \sqrt{ 1- \frac{n_\mathrm{g}^2}{n_\mathrm{EIT}\left(\delta\right)^2}\sin\vartheta}\label{Angle_2},
\end{eqnarray}
and the wave vector takes the form
\begin{equation}
 k\left(\delta\right)= -\mathrm{i}\frac{\delta}{c} \sqrt{n_\mathrm{m}^2-n_\mathrm{g}^2\sin^2\vartheta}.
\end{equation}
In all above equations, $n_{\mathrm{g,m,EIT}}$ stand for the refractive indices of the glass, metallic and EIT media, respectively. Note that atomic medium susceptibility affects both $n_{\mathrm{EIT}}$ and $\vartheta_{\mathrm{EIT}}$, which makes a qualitative difference between equations (\ref{Angle_1}) and (\ref{Angle_2}) and strongly influences the reflection coefficient $R$, allowing for tunability. As we have explicitly written, we neglect the dispersion of glass and metal and only include the frequency-dependence of the EIT properties,
since the effects here discussed apply to narrow GHz spectral ranges, comparable with atomic linewidths.

In the following section we analyse a realistic system in the context of excitation of narrow-band surface plasmons and their ultrasensitive tunability.

\section{Scattering coefficient} \label{sec:scattering}
\begin{figure*}[ht!]
\begin{centering}
a)\includegraphics[width=8.37cm,keepaspectratio]{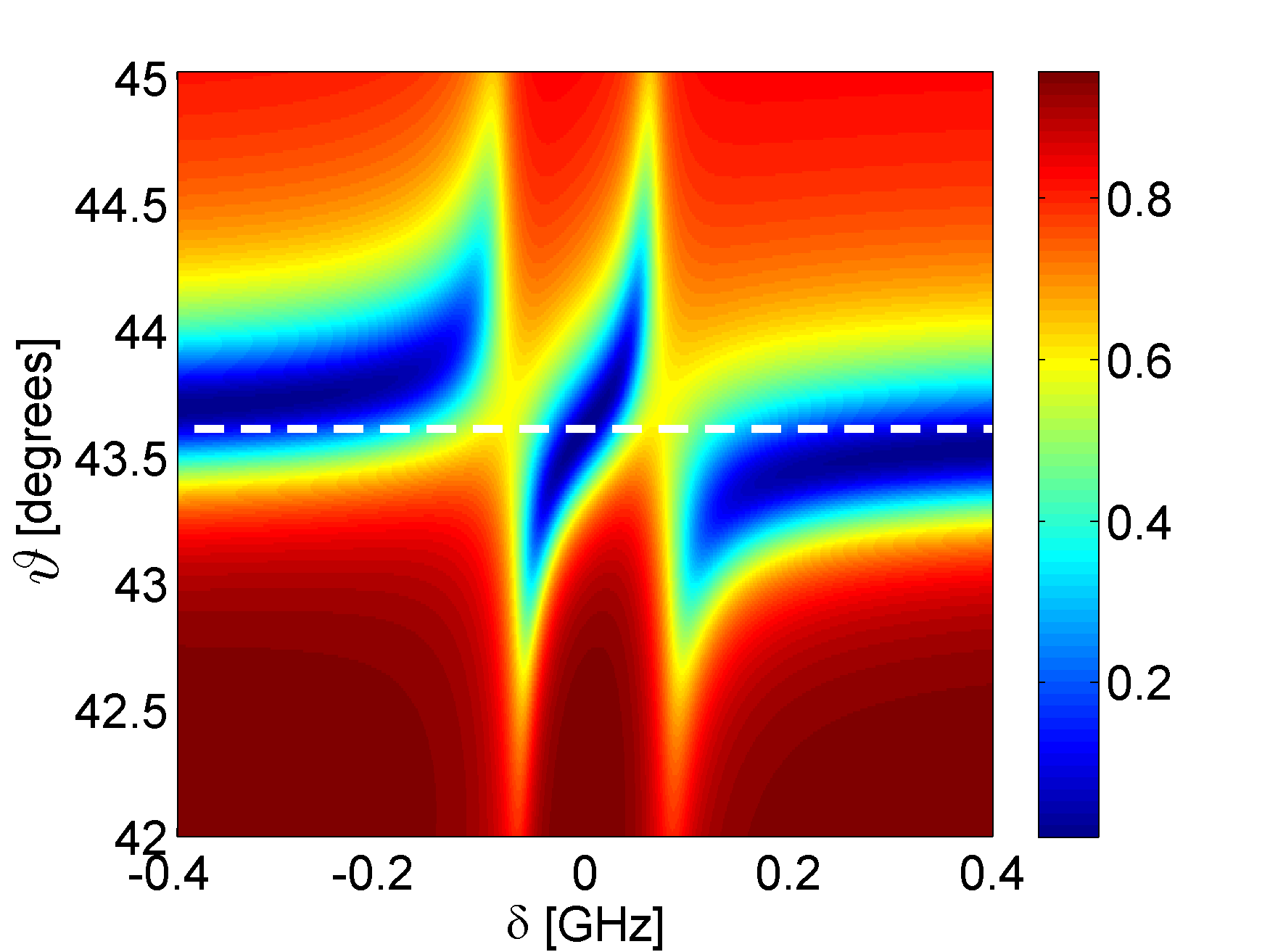}
b)\includegraphics[width=8.37cm,keepaspectratio]{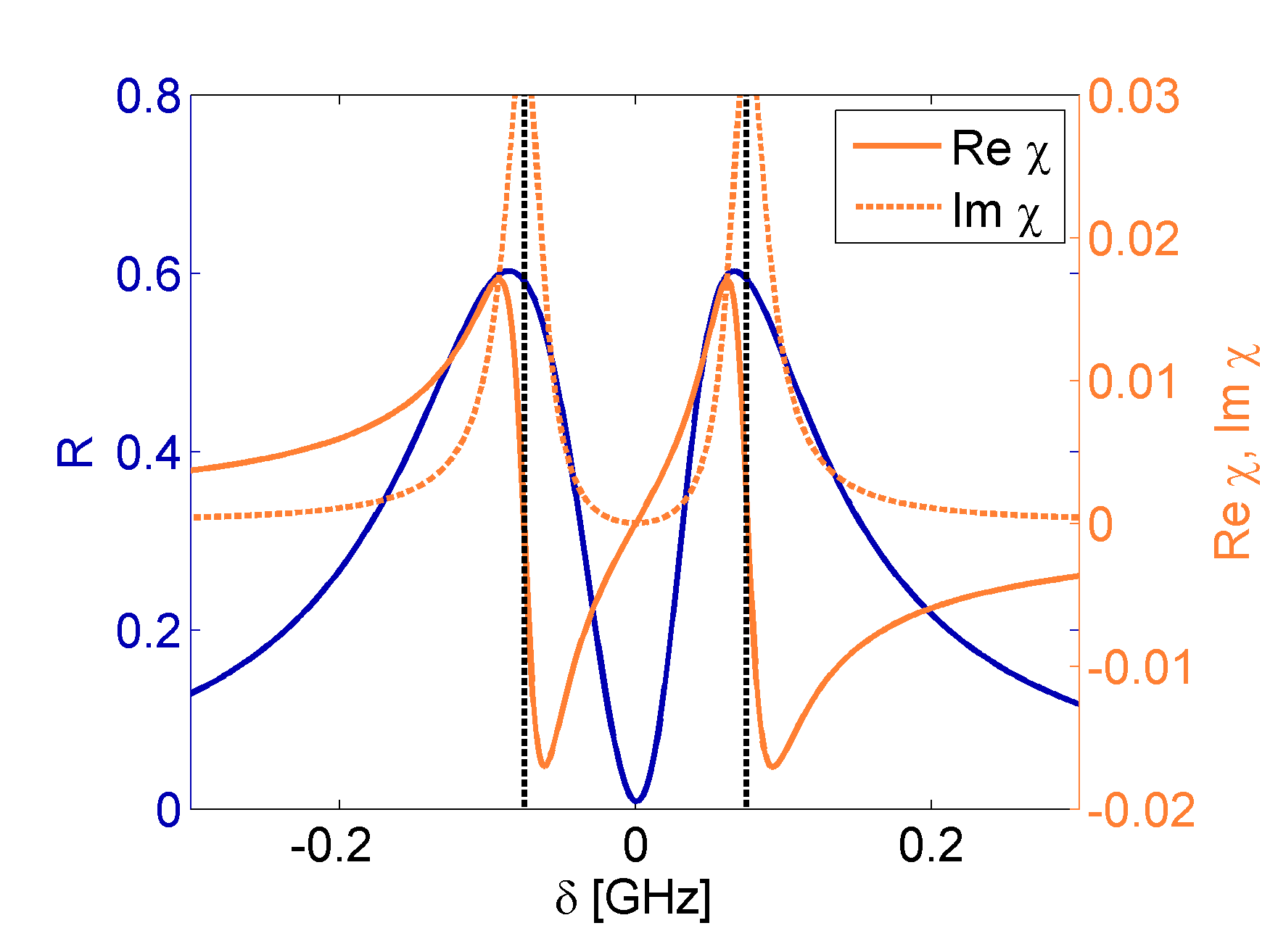}\\
c)\includegraphics[width=8.37cm,keepaspectratio]{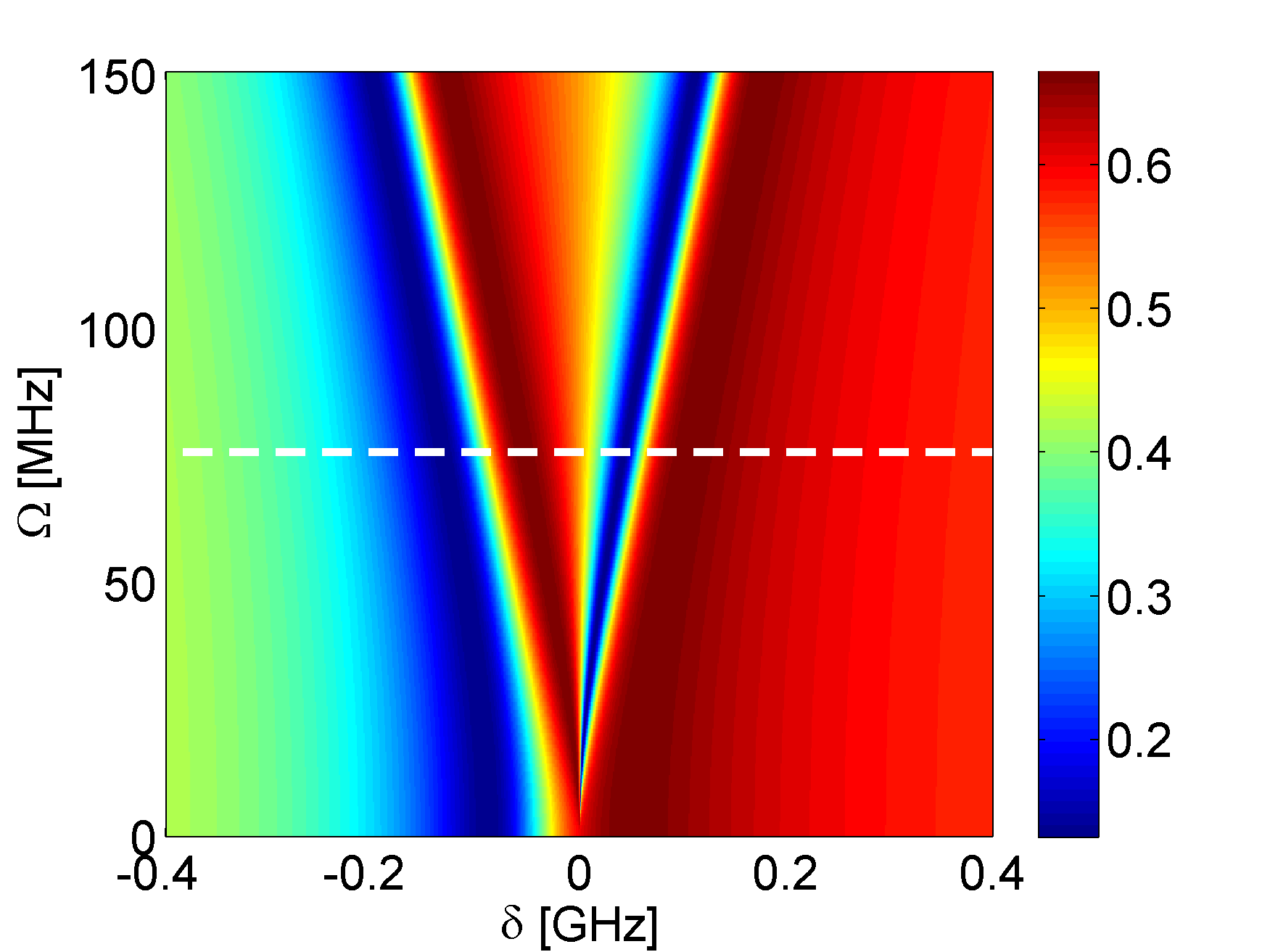}
d)\includegraphics[width=8.37cm,keepaspectratio]{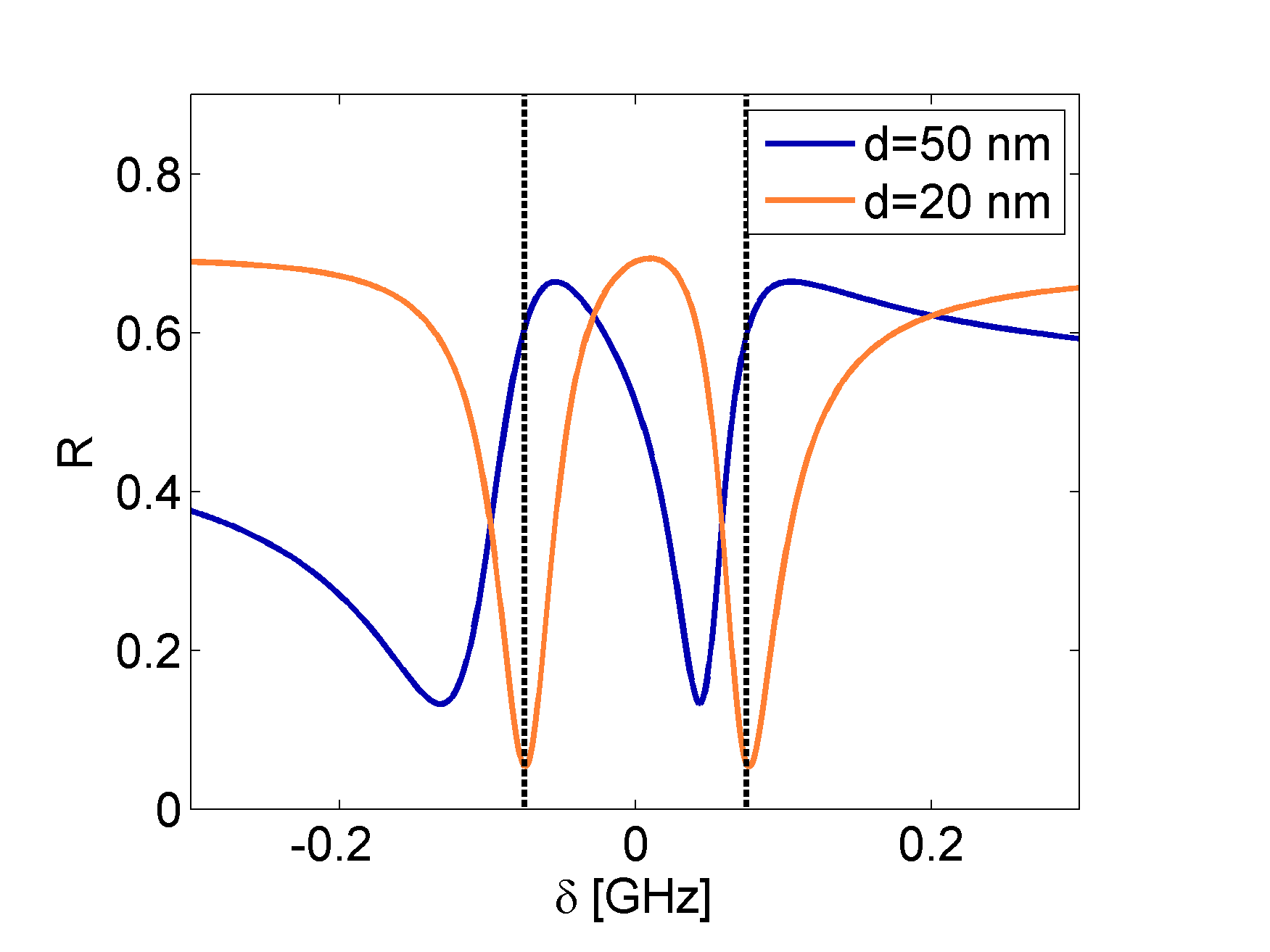}\\
e)\includegraphics[width=8.37cm,keepaspectratio]{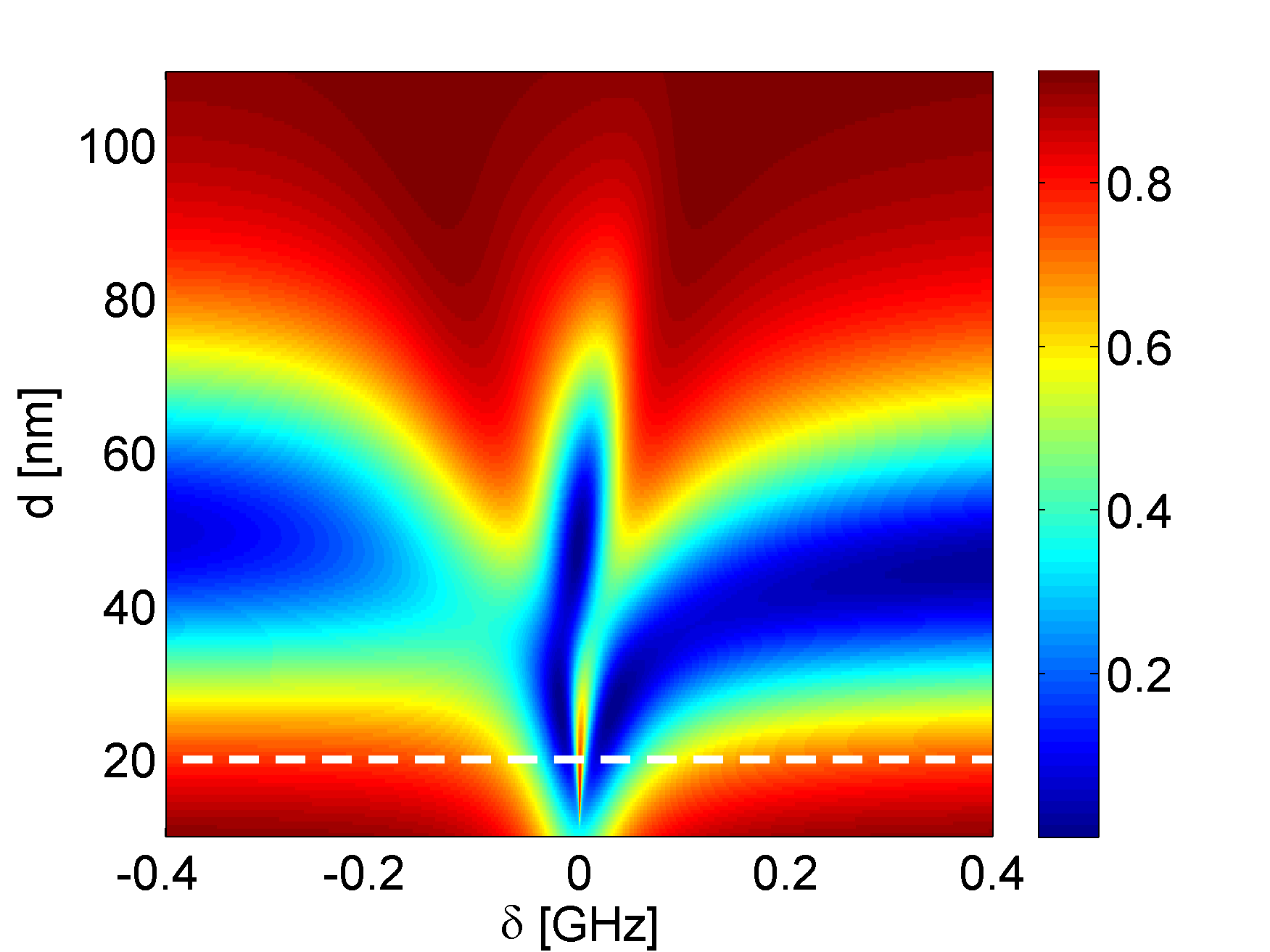}
f)\includegraphics[width=8.37cm,keepaspectratio]{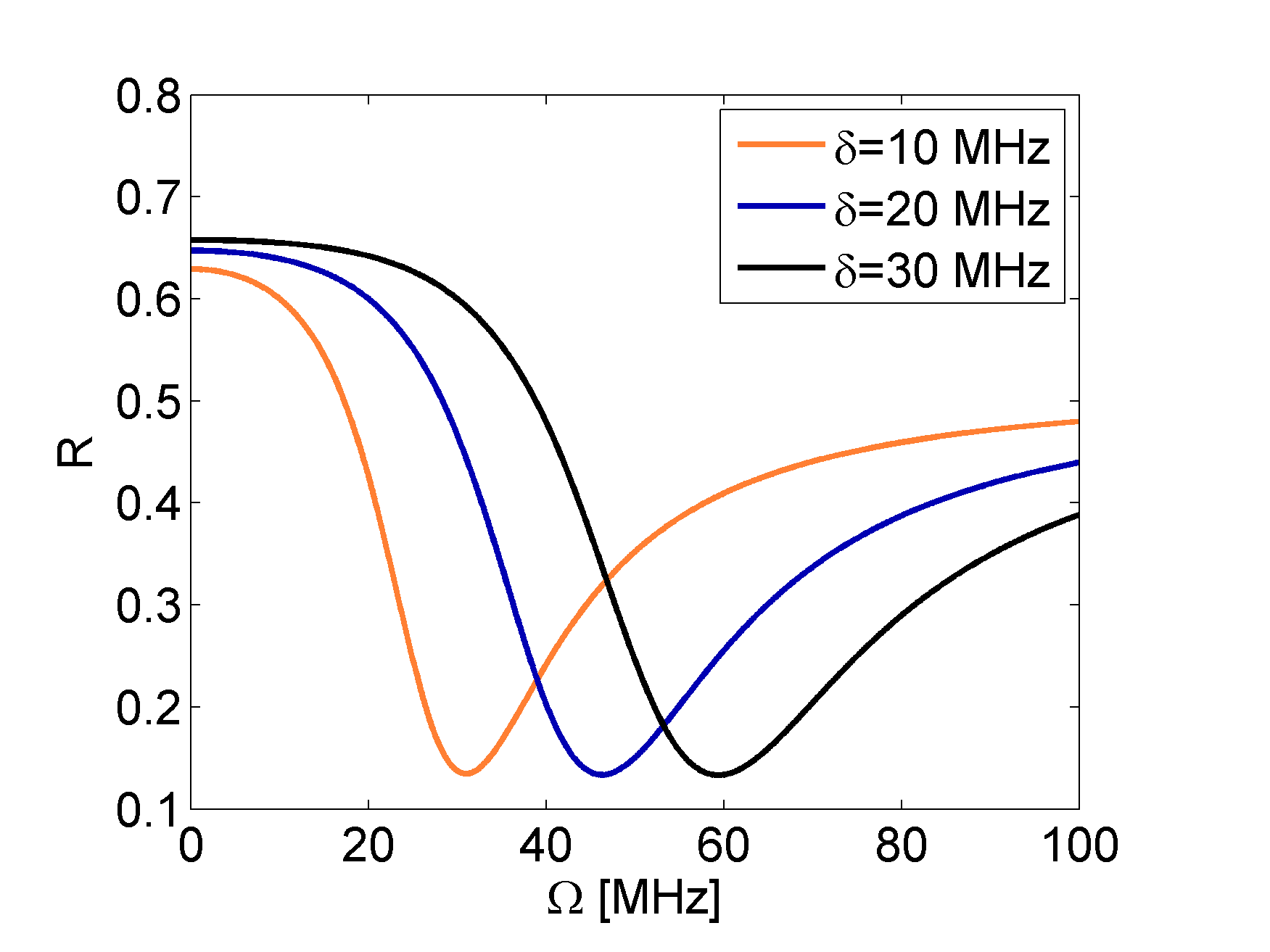}
\end{centering}
\caption{\label{fig:plasmons}Scattering spectra against various parameters of the investigated setup. Panels b) and d) correspond to cross sections of color plots a), c) and e), at parameter values marked with white dashed lines. 
a) Scattering spectra as functions of the illumination angle $\vartheta$, equal to the detection angle, for a fixed control field $\Omega = 75.7$ MHz, and for the silver layer thickness $d = 50$ nm. The white dashed line marks an optimal illumination angle, for which a resonance dip at the center of the EIT transparency window indicates a creation of a narrow-band SPP. 
b) Reflection spectrum at the optimal angle, compared with the susceptibility of EIT medium. Clearly, the SPP dip is located inside the transparency window. 
c) Scattering spectra as a function of the control field Rabi frequency $\Omega$, for the illumination angle $\vartheta=44^o$, showing two optically controlled plasmon resonances. 
d) Reflection coefficient for the fixed control field $\Omega = 75.7$ MHz and two thicknesses, corresponding to the cross sections of plots a) and c). The result for $d = 20$ nm shows clear dependence of the reflection coefficient on the EIT medium absorption.
e) Scattering spectra at an optimal illumination angle $\vartheta = 43.61^o$ and a fixed $\Omega = 75.7$ MHz, in dependence of the silver film thickness $d$. 
f) Reflection coefficient for three fixed probe detunings, as a function of Rabi frequency $\Omega$.}
\end{figure*}

For calculations, we consider silver films of thickness up to \mbox{$d < 100$ nm},
atop a glass surface of a refractive index $n_{g} = 1.5$.
The illumination beam is modelled as a plane wave incident at the metallic surface at a narrow range of illumination angles around the optimal value from the glass side. The frequency of the beam is chosen around the EIT resonance, i.e. where the most interesting effects occur. 
For this particular narrow range, the dispersion of silver can be neglected, and its permittivity fixed at
{ $\epsilon_\mathrm{m} = -13.3 + 0.883\mathrm{i}$ } \cite{Srebro}.
 
We begin with an analysis of the scattering spectra obtained for a layer of $50$ nm thickness and a control field $\Omega = 75.7$ MHz in a range of illumination angles $\vartheta \in (42^o,45^o)$. 
As follows from Fig.~\ref{fig:plasmons} a), at an optimal incidence angle $\vartheta_\mathrm{opt} \approx 43.6^o$, marked with a white dashed line, a pronounced scattering dip occurs at the center of the EIT transparency window.  
This indicates a creation of an SPP that propagates along the metal-dielectric interface (please see also Figs.~\ref{fig:setup} a) and~\ref{fig:plasmon}, in which such SPPs are captured). 
Please note that the bandwidth of the SPP excitation is remarkably narrow, i.e. below $100$ MHz. As shown in Fig.~\ref{fig:plasmons} b), the resonance conditions are met for $\mathrm{Re}~\chi=0$. Even a very small change of the medium susceptibility, altering the refraction index of EIT medium given by Eq.~\ref{eq:n}, is sufficient to impede the formation of SPPs and dramatically increase the reflection coefficient of the system. The maxima of reflection are close to the regions of peak absorption, marked by black, dashed lines. At a slightly tilted angle $\vartheta \in (43.2^o,44.2^o)$, the position of the SPP resonance in Fig.~\ref{fig:plasmons} a) is shifted, and determined by the control field $\Omega$. At the sides, an excitation of an SPP could be blocked at two narrow frequency ranges around $\pm\Omega$, i.e. at positions tunable with the control field $\Omega$. Such possibility to control the width of SPP resonance could be exploited for classical and quantum information processing with plasmonics. 

The tunability is confirmed in Fig.~\ref{fig:plasmons} c), obtained for $\vartheta = 44^o$. In this case, two SPP resonances appear: a relatively broad one that corresponds to $\delta<0$, and a narrow one for $\delta>0$, located inside the transparency window. Their central positions are shifted proportionally to the strength of the control field generating EIT. The line asymmetry is clearly visible in Fig.~\ref{fig:plasmons} d). The comparison with panel a) reveals that for the chosen illumination angle, the SPP are generated for $\mathrm{Re}~\chi=0.01$, which occurs only near the EIT resonances. Notably, for $d=50$ nm, the medium absorption in these regions is sufficiently small not to affect the formation of SPPs.

The dependence of the scattering spectra on the silver film thickness $d$ is presented in Fig.~\ref{fig:plasmons} e). For $d=50$ nm the narrow SPP resonance discussed above appears. 
Similar structures can be identified for smaller thicknesses $d\sim 20$ nm. 
As we have verified, they do not correspond to creation of SPPs, but result from absorption in the EIT media, dominant for so thin metallic layers. 
In this case, the incident wave can penetrate deeper into the EIT medium, where it may be absorbed at frequencies corresponding to the edges of the transparency window. This is clearly seen in Fig.~\ref{fig:plasmons} d), where the reflection spectrum for $d=20$ nm is presented. The narrow dips of reflection coefficient correspond to absorption maxima.

Finally, the crucial role of the EIT phenomenon for the narrow-band SPP creation is even stronger emphasized in Fig.~\ref{fig:plasmons} f).
The resonance condition can only be fulfilled due to a modification of the dispersion relation of the medium by the EIT,
which is only achieved for a narrow range of the control field strengths $\Omega$, with a radical cut-off once the condition is broken.
Please note that such behaviour repeats for a range of illumination-beam frequencies (detunings), and only requires adjusting the control field.
Such sensitivity to the field strength suggests a possibility of dynamic tuning, with applications for switchable devices for SPPs excitation and processing.

To summarize, the SPP tunability can be achieved through a dynamical adjustment of the control field:
\begin{itemize}
 \item The Rabi frequency $\Omega$ directly impacts the spectral width of transparency window, and in consequence of the generated plasmons (Fig. \ref{fig:plasmons} a).
 \item A combination with a tilted illumination angle $\vartheta$ provides means to shift to a certain extent also the very spectral position of plasmonic excitations (Fig. \ref{fig:plasmons} c).
 \item Similarly, such an overall shift can be achieved through a frequency detuning $\delta_{ac}$ of the control field (Eq. \ref{eq:chi}). 
\end{itemize}

The discussed scattering spectra have been obtained analytically.
In the Appendix, we cross-check these results against rigorous Finite-Difference-Time-Domain (FDTD) simulations.

\section*{Conclusions} \label{sec:conclusions}
In conclusion, we have demonstrated that EIT surroundings might be used to overcome the two major limitations of plasmonics:
(1) the lack of tunability, and (2) the broad SPP resonances, detrimental for quantum apllications. The key idea is to introduce the optical tunability of EIT to the field of plasmonics. 
The creation of SPPs is possible for an optimal range of metal layer thicknesses,
but can be sensitively tuned with the illumination angle and control field strength. We have shown a possibility of excitation of narrow-band SPPs within the EIT transparency window,  
and examined the optimal conditions for such action.  
The crucial role of the EIT effect is manifested in a sharp cut-off at the control-field dependence of the SPP creation window.
Narrow-band SPPs might be useful, e.g., for high-precision metrology,
while the sensitive tuning might open an entirely new avenue in nanoengineering.
The potential applications include signal processing through photonic logic gates or switches.
An extension to dynamic tuning might empower an optical response of nanostructures modified in time.

\section*{Appendix: Numerical simulations} \label{sec:fdtd}
The Finite-Difference-Time-Domain method is a valuable tool in analysis of plasmonic systems \cite{Okada}, and it can be suitably extended to simulate electromagnetically induced transparency \cite{my_EIT}. The simulation setup is shown in Fig. \ref{fig:setup} a). The calculation domain is two - dimensional. The electric field $\vec E=[E_x,E_y,0]$ has two components in the plane of incidence, and magnetic field $\vec H = [0,0,H_z]$ is perpendicular to the plane.
For the metal layer, we have used the Auxillary Differential Equations (ADE) method based on a Drude model. The particular implementation and parameters of the model were adapted from Ref. \cite{Okada}. For the considered frequency of D2 Sodium line, the permittivity of the silver layer is $\epsilon=\epsilon_r + i\epsilon_i = -13.3+0.883i$ and is described by Drude model
\begin{equation}
\epsilon(\omega)=1-\frac{\omega_p^2}{\omega^2 -i\gamma\omega}
\end{equation}
where
\begin{eqnarray}
\omega_p = \omega\sqrt{\frac{(1-\epsilon_r^2)+\epsilon_i^2}{1-\epsilon_r}},\nonumber\\
\gamma = \omega\frac{\epsilon_i}{1-\epsilon_r}.
\end{eqnarray}
The above parameters are used to solve the time domain equation for medium polarization $P$
\begin{equation}
\ddot{P} - \gamma \dot{P} = \epsilon_0\omega_p^2 E
\end{equation}
which, in turn, is used in the FDTD relations
\begin{eqnarray}
\frac{\partial H_z}{\partial t}=\frac{1}{\mu_0}\left(\frac{\partial E_x}{\partial y} - \frac{\partial E_y}{\partial x}\right), \nonumber\\
\frac{\partial E_x}{\partial t}=\frac{1}{\epsilon_0}\left(\frac{\partial H_z}{\partial y}-\frac{\partial P_x}{\partial t}\right),\nonumber\\
\frac{\partial E_y}{\partial t}=\frac{1}{\epsilon_0}\left(-\frac{\partial H_z}{\partial x}-\frac{\partial P_y}{\partial t}\right).
\end{eqnarray}

The metal layer has a thickness of $2\Delta x$, where $\Delta x$ is the minimum spatial step of the simulation. This corresponds to $\lambda/5$ for the highest considered frequency.
\begin{figure}[ht!]
\includegraphics[width=8.2cm,keepaspectratio]{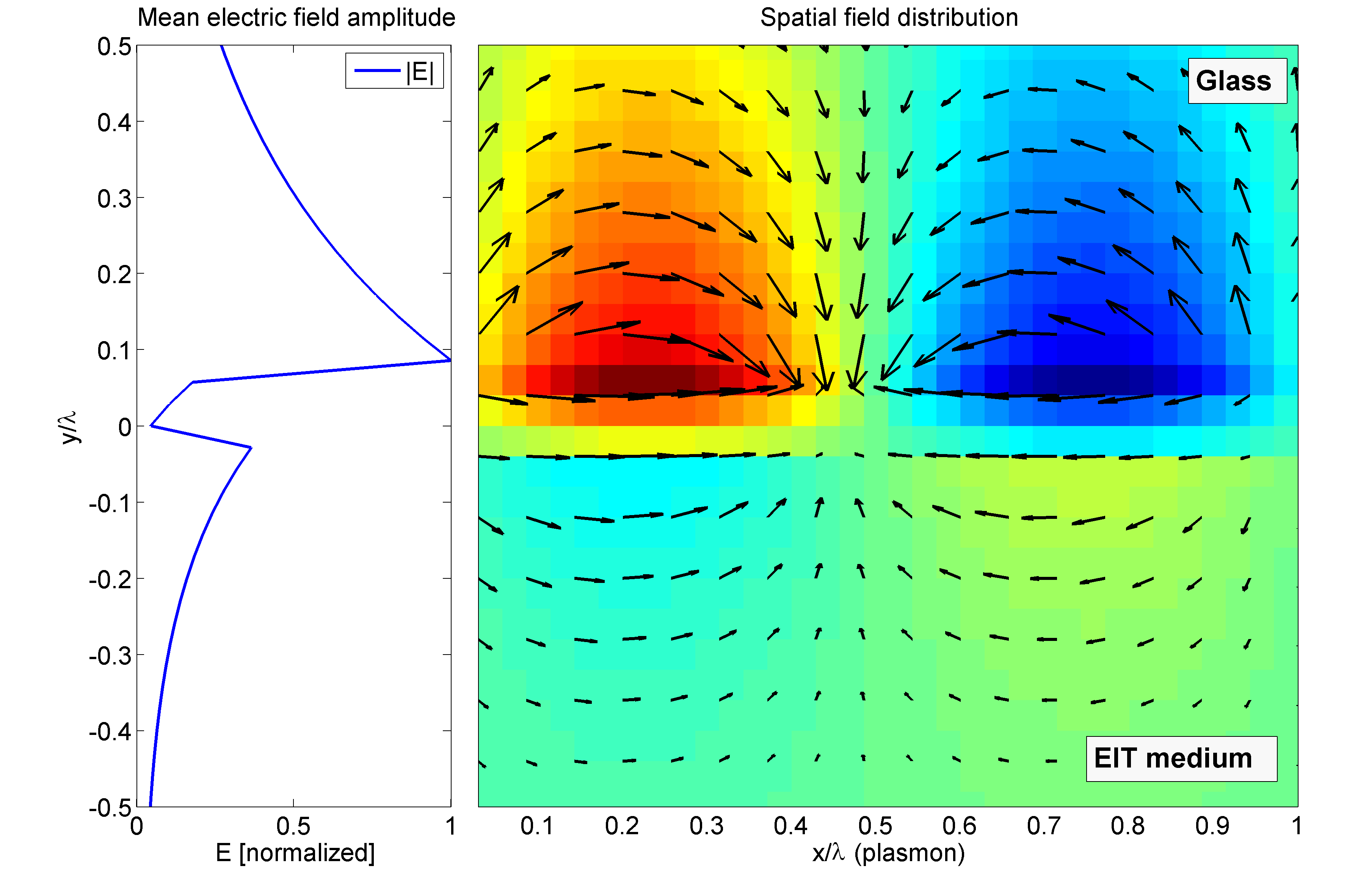}
\caption{\label{fig:plasmon}Right panel: magnetic field amplitude (color) and electric field vectors (black arrows) of the SPP on a conductive sheet, simulated with FDTD. Left panel: The mean electric field amplitude along the direction perpendicular to the metal surface.}
\end{figure}

As a first validation, we have confirmed that the modelled system supports the creation and propagation of SPPs. The results are shown in Fig. \ref{fig:plasmon}. One can see that the field is confined to the metal surface, decaying exponentially away from it. As expected, the electric field inside the metal is very low. Both the wavelength and phase velocity of the SPP is consistent with the dispersion relation in Eq. (\ref{eq:k}). 

\begin{figure}[ht!]
\begin{centering}
\includegraphics[width=8.4cm,keepaspectratio]{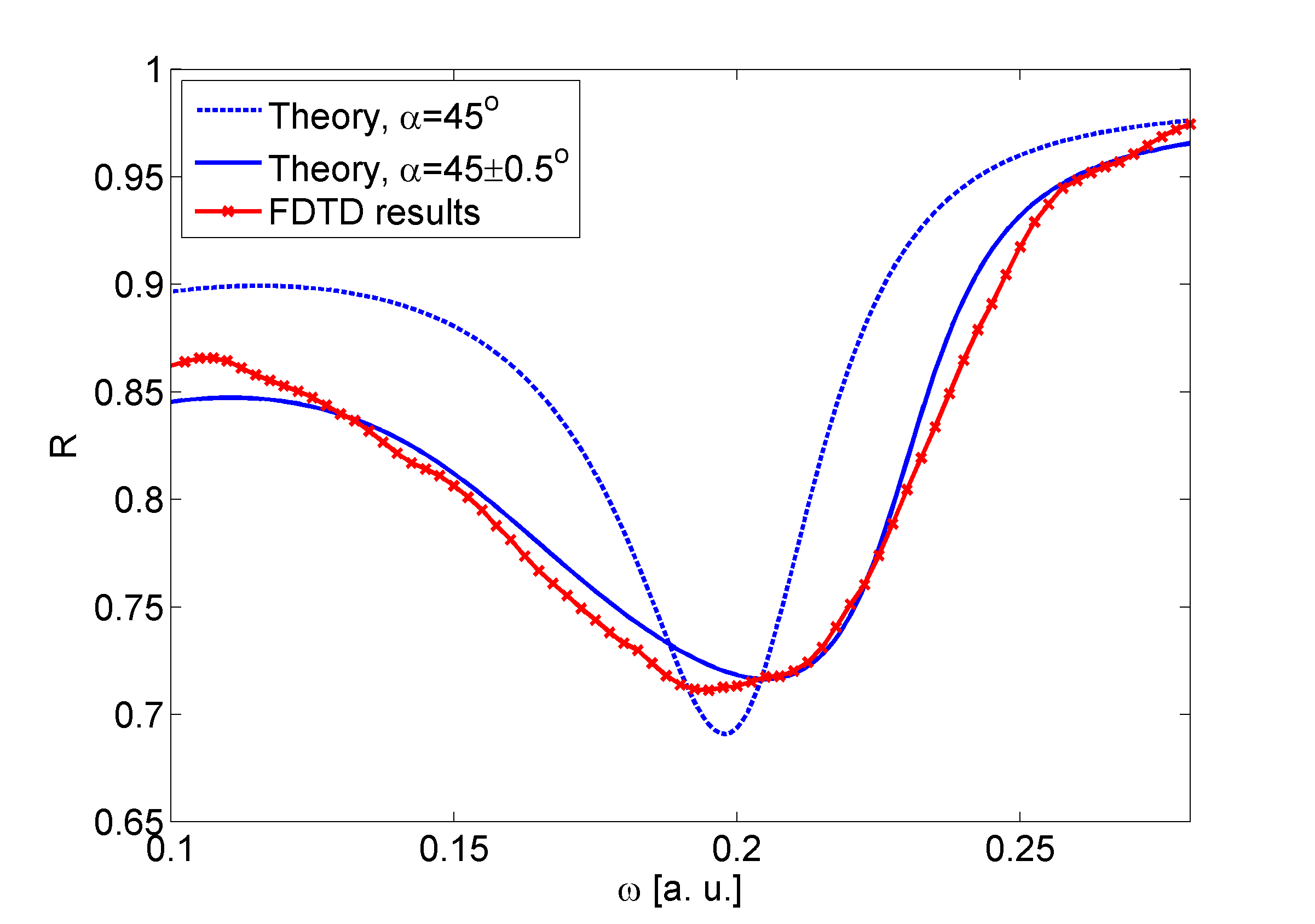}\\
\includegraphics[width=8.4cm,keepaspectratio]{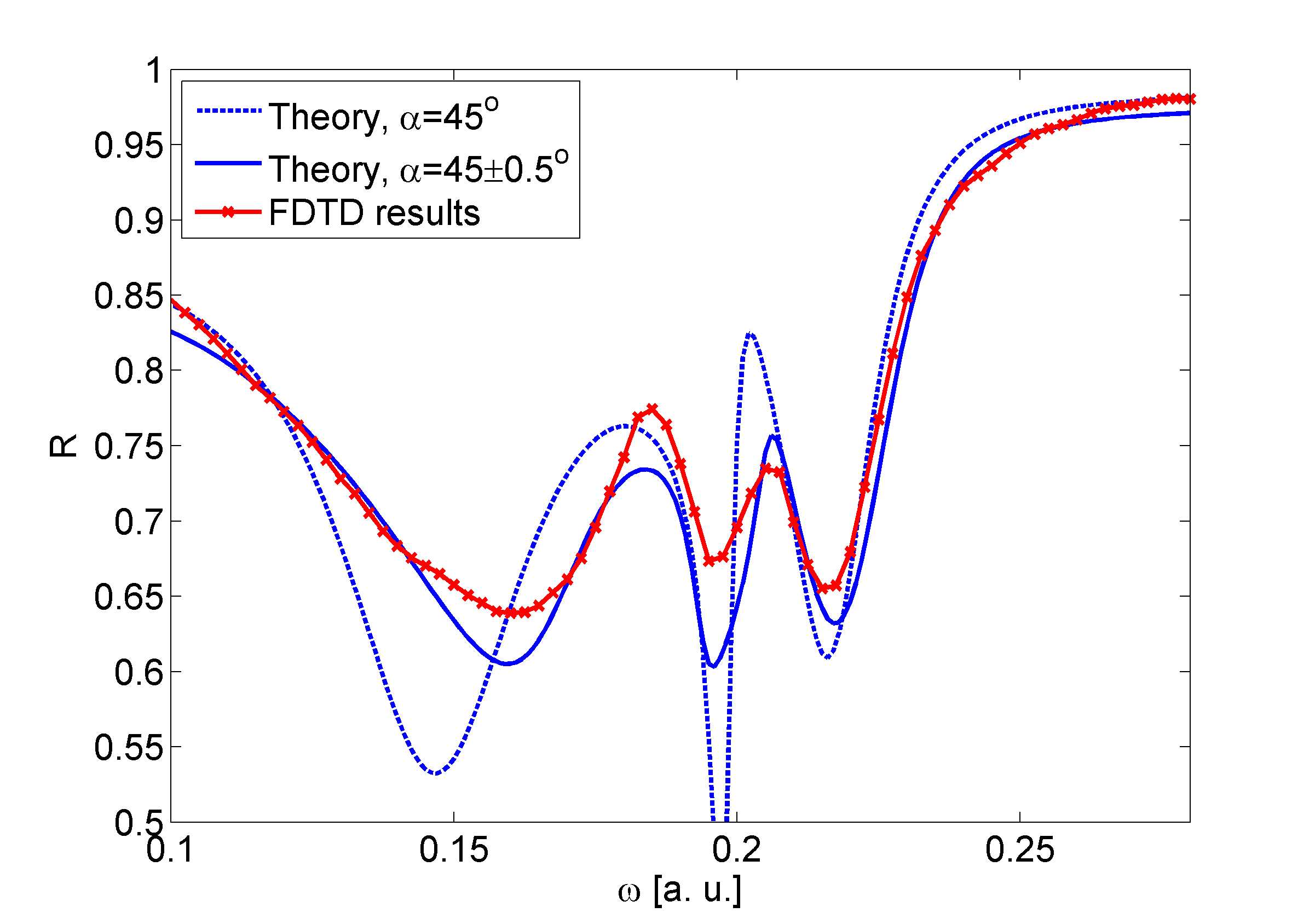}
\par
\end{centering}
\caption{\label{fig:numer}a) FDTD simulation results for SPP resonance for glass-metal-vacuum system, at the optimal incidence angle $\vartheta = 45^o$ and a range of frequencies. The analytical solution is averaged over a range of incidence angles. b) Results for plasmon resonance with EIT medium.}
\end{figure}

Next, we have simulated an SPP resonance in the case where the surrounding medium is vacuum. The results were compared with the analytical solution given by Eq. (\ref{Rcoeff}). Due to the constrained simulation time, a relatively wide frequency range has been used, so that the dispersion of the metal was not negligible and has been taken into account in the analytical solution. The dispersive properties of the metal were adjusted to yield the SPP resonance at the incidence angle $\vartheta=45^o$. Due to the nature of FDTD simulation, the modelled region of space is limited. As seen in Fig. \ref{fig:setup} a), the incident beam is relatively narrow, with the width comparable to wavelength. This means that diffraction plays a significant role in the system, putting uncertainty in the incidence angle. The obtained reflection spectrum is shown in Fig.~\ref{fig:numer} a), and exhibits a single, wide dip corresponding to the surface plasmon resonance. One can see that numerical results closely match the analytical solution when suitable angle correction is included.
As a next step, we have simulated a system consisting of glass, metal layer and atomic medium. The numerical model of electromagnetically induced transparency is based on standard approach used in simulation of EIT metamaterials \cite{my_EIT,my_czer,Zhang}. In comparison to atomic system, the transparency window is very wide (of the order of 0.1 $\omega$), covering significant fraction of the SPP resonance. Likewise, the damping constant $\gamma$ is proportionally bigger, increasing the numerical stability. Apart from these differences, the simulation encompasses all the key characteristics of the system under consideration. We have chosen the case where the EIT window is located in the middle of the SPP dip. The results are shown in Fig.~\ref{fig:numer} b). 
All of the characteristic features of the solution presented in Fig.~\ref{fig:plasmons} b) are present; the changes of the permittivity due to the EIT impede the plasmon formation, generating two reflection peaks in the spectrum, separated by a narrow dip located inside the transparency window. Again, there is a good agreement between FDTD results and angle-averaged analytical result.

\end{document}